\documentstyle[psfig,preprint,aps]{revtex}

\newcommand{\sla}{\!\!\!\!/ \,}

\begin{document}
\hoffset-1cm
\draft

\title{ Ward Identities in Non-equilibrium  QED\footnote{Supported by BMBF, GSI Darmstadt, DFG and NSFC} }

\author{M.E. Carrington$^1$, Hou Defu$^2$, and Markus H.
Thoma$^3$\footnote{Heisenberg fellow}}
\address{$^1$Physics Department, Brandon University, Brandon Manitoba R7A 6A9, Canada
   }
\address{$^2$Institut f\"ur Theoretische Physik, Universit\"at Regensburg,
   D-93040 Regensburg, Germany\\
 and Institute of Particle Physics, Huazhong Normal University, 430070
Wuhan,China}
\address{$^3$Institut f\"ur Theoretische Physik, Universit\"at Giessen,\\
   D-35392 Giessen, Germany}

\date{\today}

\maketitle

\begin{abstract}

We verify the QED Ward identity for the two- and three -point functions at non-equilibrium in 
the HTL limit.  We use the  Keldysh formalism of real time  finite temperature field theory.  We obtain an identity of the same form as the Ward identity for a set of one loop self-energy and one loop three-point vertex diagrams which are constructed from HTL effective propagators and vertices.  \end{abstract} 
%\pacs{PACS numbers: 11.10Wx, 11.15Tk, 11.55Fv}
\narrowtext
\newpage
 %%%%%%%%%%%%%%%%%%%%%%%%%%%%%%%%%%%%%%%%%%%%%%%%%%%%%%%%%%%%%%%%%%
\section{Introduction}

There has been much progress in our understanding of field theory at finite
temperature, however, the  perturbation theory of gauge
theories is still plagued with serious problems: some physical
quantities, when computed perturbatively, suffer from gauge dependence and
infrared divergences. Many of these problems are resolved by the use of the Braaten and 
Pisarski resummation scheme \cite{ref1,ref2}. In many theories (for example QED and QCD)  there are amplitudes with soft ($\sim gT$) external lines that have loop `corrections' 
that are of the same order in perturbation theory as the tree diagrams.  These additional 
contributions are one-loop diagrams in which the loop momentum is hard ($\sim T$).   In such a theory, in order to systematically calculate amplitudes with soft lines, it is necessary to do perturbation theory within an effective theory that is obtained from the original theory by resumming `hard thermal loops' (HTLs).  For soft lines, effective propagators 
are used in which the bare propagator is corrected by a self-energy insertion given by 
the one loop graph with the momentum integral restricted to the hard momentum range (which is 
the region that dominates the integral - see Fig. 1).  For vertices in which all external legs
are soft, the effective vertex is given by the bare vertex plus the one-loop vertex, where 
the momentum integral is restricted to the hard momentum range (Fig. 2).  These are the hard 
thermal loop effective propagators and vertices.  If a line is 
hard, or if a vertex contains at least one external line that is hard, loop corrections 
are suppressed by at least one power of the coupling constant, and therefore bare 
propagators and vertices can be used to leading order.  In particular, it is consistent 
to use bare propagators and vertices to calculate the HTLs themselves.  In addition, the HTLs are gauge invariant \cite{ref1,taylor1,BP92}.  Thus, the Braaten and Pisarski resummation 
scheme provides a self-consistent, gauge invariant way to do perturbative calculations for 
amplitudes with soft external momenta.  

The HTL effective perturbation theory was derived within the
imaginary formalism  and attention was  mainly focused on equilibrium field theory.  
However, realistic physical systems are frequently out of equilibrium.\footnote{Physically 
interesting examples 
are the central region of high energy heavy ion collisions, and the phase transition 
of the electroweak theory in which the Higgs condensate evaporates.}  We want to consider how microscopic processes change when they take place in a background distribution that is not at equilibrium.  It is assumed that the time scale of these microscopic processes is much less than the time scale with which the background relaxes towards equilibrium.   

Non-equilibrium 
calculations must be carried out in the real time formalism \cite{ref6,ref7}.  One of the 
traditional difficulties associated with the real time formalism is the doubling of 
degrees of freedom.  In equilibrium, it is easy  to see that these additional degrees of freedom are necessary to affect the cancellation of pinch  singularities.  
Out of equilibrium, the situation is not as straightforward.  In an earlier paper  
we have proven that pinch singularities do not occur in the non-equilibrium HTL effective 
propagator \cite{mhm}. When the imaginary part of the HTL contribution to the self energy is zero, the singularity does not occur.  When the imaginary part is non-zero, the HTL effective propagator is given by a resummation 
of pinch terms that is finite.  This result indicates that perturbative calculations are 
sensible in a non-equilibrium field theory.\footnote{Non-equilibrium HTL calculations involve the  assumption that there is some scale above which the non-equilibrium statistical distribution functions vanish sufficiently fast, and that these distribution functions are not too singular in the infra-red region.  These assumptions ensure that when calculations are done by taking loop momenta to be much larger than external momenta, the resulting HTLs give the dominant contributions at one loop.}   

In this paper we use the Keldysh formulation of real time thermal field theory \cite{ref6,ref7}.  
In the next section we explain this formalism.  
  In the third section of this paper we will 
show that the gauge invariance of HTLs in equilibrium persists when the system is out of equilibrium; we will study 2 and  3-point  HTLs in QED and verify that the Ward identities 
are obeyed out of equilibrium.  This result shows that non-equilibrium  HTL calculations 
are analogous to calculations in the equilibrium case.  
In the fourth section of this paper we consider a type of higher order effective theory.   We will consider a subset of diagrams from such a second order effective theory and show that they satisfy an identity which has the same form as the Ward identity.  
%%%%%%%%%%%%%%%%%%%%%%%%%%%%%%%%%%%%%%%%%%%%%%%%%%%%%%%%%%%%%%%%%%
\section{Keldysh Formalism}
\label{sec2}
%%%%%%%%%%%%%%%%%%%%%%%%%%%%%%%%%%%%%%%%%%%%%%%%%%%%%%%%%%%%%%%%%%

In the Keldysh formalism, the closed time path contour is used.  
This contour has two branches, ${\cal C}_1$ going from negative infinity to positive 
infinity at a distance $+i\epsilon$ above the real axis, and ${\cal C}_2$ returning from 
positive infinity to negative infinity just below the real
axis\cite{ref6,ref7}. 
\subsection{Propagator}

The propagator has four components given by \cite{ref6}
 \begin{eqnarray}
 \label{eq: compD}
   G_{11}(x-y) &=& -i\langle T(\phi(x) \phi(y))\rangle \, , \nonumber\\ 
   G_{12}(x-y) &=& \mp i\langle \phi(y) \phi(x) \rangle \, , \nonumber\\
   G_{21}(x-y) &=& -i\langle \phi(x) \phi(y)\rangle \, , \nonumber\\
   G_{22}(x-y) &=& -i\langle\tilde{T}(\phi(x)\phi(y))\rangle \, , 
 \end{eqnarray}
where $T$ is the usual time ordering operator, $\tilde{T}$ is the 
antichronological time ordering operator, and the subscripts $\{1,2\}$ refer to the 
contours along which the fields take values. $G$ is either a scalar propagator (D)  
or a spinor propagator (S), with the upper sign taken for bosons and the lower sign 
for fermions.  As a consequence of the identity $\theta(x) + \theta(-x) =1$, these 
four components 
satisfy,
 \begin{equation}
 \label{3}
   G_{11} - G_{12} - G_{21} + G_{22} = 0. 
 \end{equation} 

It is more useful to write the propagator in terms of the three functions
 \begin{eqnarray} \label{3a}
   G_R &=& G_{11} - G_{12} \, , \nonumber\\
   G_A &=& G_{11} - G_{21} \, , \nonumber\\ 
   G_F &=& G_{11} + G_{22} \, .
 \end{eqnarray}
$G_R$ and $G_A$ are the usual retarded and advanced propagators.
In equilibrium these propagators are related: 
 \begin{equation}
 \label{5}
   G_F(P) = N(P) \Bigl(G_R(P) - G_A(P)\Bigr). 
 \label{eq: KMSD}
 \end{equation}
Throughout this paper we define:  for bosons 
$N(P) \equiv N_B(P) = [1+2n_B(p_0)]\mbox{sgn} (p_0)$ with $n_B(p_0)$ the thermal 
Bose-Einstein distribution; for fermions 
$N(P) \equiv N_F(P) = [1-2n_F(p_0)]\mbox{sgn} (p_0)$ with $n_F(p_0)$ the thermal Fermi-Dirac 
distribution.  Out of equilibrium,~(\ref{eq: KMSD}) holds for bare propagators when the 
thermal distribution functions are replaced by the appropriate Wigner functions: $n_B 
\rightarrow f_B$ and $n_F \rightarrow f_F$. \cite{mhm}.

The self energies, or 1PI 2-point functions are obtained from the propagators by truncating 
external legs.  The retarded and advanced self energies are defined as, 
\begin{eqnarray} \Sigma_R &=& \frac{1}{2}[\Sigma_{11} + \Sigma_{12}-\Sigma_{21} -\Sigma_{22}],
\nonumber \\
\Sigma_A &=& \frac{1}{2}[\Sigma_{11} + \Sigma_{21} - \Sigma_{12} -\Sigma_{22}],
\label{sigma}
\end{eqnarray}
where we have used the identity,
\begin{eqnarray}
\Sigma_{11} + \Sigma_{12} + \Sigma_{21} + \Sigma_{22} =0.
\end{eqnarray}
  The symmetric self-energy is given by,
\begin{eqnarray}
\Sigma_F &=& \frac{1}{2}[\Sigma_{11} + \Sigma_{22} - \Sigma_{12} -\Sigma_{21}]
\label{sigmaf}
\end{eqnarray}
and satisfies, in equilibrium, 
\begin{eqnarray}
\Sigma_F(P) = N(P)(\Sigma_R(P) - \Sigma_A(P)).
\end{eqnarray}

In real time, the simplest way to do calculations is to use tensor forms 
for the propagators~(\ref{decompD1}).  These forms are obtained by inverting~(\ref{3}) 
and~(\ref{3a}), 
 \begin{eqnarray}
 \label{7}
   G_{11} &=& \frac{1}{2} (G_F + G_A + G_R) \, , \nonumber\\
   G_{12} &=& \frac{1}{2} (G_F  +G_A - G_R) \, , \nonumber\\
  G_{21} &=& \frac{1}{2} (G_F  -G_A + G_R) \, , \nonumber\\
 G_{22} &=& \frac{1}{2} (G_F  -G_A - G_R) \, , 
 \end{eqnarray}
and rewriting as \cite{PeterH}
 \begin{equation}
 \label{decompD1}
   2\,G = G_R {1\choose 1}{1\choose -1} 
        + G_A {1\choose -1}{1\choose 1} 
        + G_F {1\choose 1}{1\choose 1},
 \end{equation} 
where the outer product of the column vectors is to be taken. 

 We establish the following notation. Throughout this paper, photons always carry the 
momentum $S$.  We write $S^2\equiv s_0^2-s^2$, $s\equiv |{\bf s}|$, and work in the covariant gauge with the Feynman choice of the gauge parameter so that 
the photon propagator is \begin{eqnarray}
-i D_{\mu\nu}(S) = -ig_{\mu\nu} D(S), 
\end{eqnarray}
where $D(S)$ is given by~(\ref{decompD1}) with 
\begin{eqnarray}
D_R(S) = \frac{1}{(s_0+i\epsilon)^2-s^2}, \,\,\,\,\,\,\,\,\,\,D_A(S) = 
\frac{1}{(s_0 - i\epsilon)^2-s^2}
\label{RA}
\end{eqnarray}
and $D_F(S)$ is obtained from~(\ref{eq: KMSD}) as $N_B(S)(D_R(S) - D_A(S))$. 

Fermions carry momentum $P_1 = P+S$ or $P_3 = Q+S$.  The fermion propagator has the form, 
\begin{eqnarray}
iS_{\alpha\beta}(P_1) = i(P_1 \sla)_{\alpha\beta}D(P_1),
\end{eqnarray}
We can use~(\ref{decompD1}) to write the fermion propagator as the combination of three terms where, 
\begin{eqnarray}
i(S_R)_{\alpha\beta}(P_1) = i(P_1 \sla)_{\alpha\beta}\,D_R(P_1),\,\,\,{\rm etc.}
\end{eqnarray} 
Eq.~(\ref{eq: KMSD}) gives $D_F(P_1) = N_F(P_1)(D_R(P_1) - D_A(P_1))$ or $S_F(P_1) = N_F(P_1)(S_R(P_1) - S_A(P_1))$.   We will use the shorthand notation $D_R(P_1) = r_1,$ $D_R(S) = r_S$ and $S_R(P_1) = \tilde{r}_1$ ${\rm etc.}$

\subsection{Three-Point Vertex Function}

In the real time formalism the three-point function has $2^3 = 8$ 
components. We denote connected three-point functions by $\Gamma^c_{abc}$ where 
$\{a,b,c = 1,2\}$.  
The eight components are not independent because of the identity
 \begin{equation} 
   \sum_{a=1}^2\sum_{b=1}^2\sum_{c=1}^2 
   (-1)^{a+b+c-3} \Gamma^c_{abc} =0,
 \label{eq: circVer}
 \end{equation}
which follows in the same way as~(\ref{3}) from $\theta(x) + 
\theta(-x) = 1$. The retarded combinations are given by 
 \begin{eqnarray}
   \Gamma^c_{R} &=& \Gamma^c_{111} - \Gamma^c_{112} 
                - \Gamma^c_{211} + \Gamma^c_{212}, 
 \nonumber\\ 
   \Gamma^c_{Ri} &=& \Gamma^c_{111} - \Gamma^c_{112} 
                   - \Gamma^c_{121} + \Gamma^c_{122}, 
 \nonumber\\ 
   \Gamma^c_{Ro} &=& \Gamma^c_{111} - \Gamma^c_{121} 
                   - \Gamma^c_{211} + \Gamma^c_{221}, 
\nonumber\\ 
   \Gamma^c_{F} &=& \Gamma^c_{111} - \Gamma^c_{121} 
                + \Gamma^c_{212} - \Gamma^c_{222},
 \nonumber\\
   \Gamma^c_{Fi} &=& \Gamma^c_{111} + \Gamma^c_{122} 
                   - \Gamma^c_{211} - \Gamma^c_{222}, 
 \nonumber\\
   \Gamma^c_{Fo} &=& \Gamma^c_{111} - \Gamma^c_{112} 
                   + \Gamma^c_{221} - \Gamma^c_{222}, 
 \nonumber\\
   \Gamma^c_{E} &=& \Gamma^c_{111} + \Gamma^c_{122} 
                + \Gamma^c_{212} + \Gamma^c_{221}.
 \label{eq: physVer}
 \end{eqnarray} 
In coordinate space we always label the first leg of the three-point 
function by $x$ and call it the ``incoming leg $(i)$", the third leg we 
label by $z$ and call it the ``outgoing leg $(o)$", and the second 
(middle) leg we label by $y$. 

For our purposes we will need the 1PI three-point functions ${\Gamma}(P_1,P_2,P_3)$ which 
are obtained from the connected vertex functions $\Gamma^c(P_1,P_2,P_3)$ by truncating 
external legs.  For example, we have,  
\begin{eqnarray}
\Gamma^c_{R} &=& i^3 a_1 r_2 a_3 {\Gamma}_{R},\nonumber \\
\Gamma^c_{Ri} &=& i^3 r_1 a_2 a_3 {\Gamma}_{Ri}, \nonumber \\
\Gamma^c_{Ro} &=& i^3 a_1 a_2 r_3 {\Gamma}_{Ro}. \label{gammac}
\end{eqnarray}

The retarded 1PI vertex functions are given by,
\begin{eqnarray}
\Gamma_R &=& \frac{1}{2}(\Gamma_{111} + \Gamma_{112} 
+\Gamma_{211} + \Gamma_{212} - \Gamma_{121} - \Gamma_{122} 
- \Gamma_{221} - \Gamma_{222}),\nonumber \\
\Gamma_{Ri} &=& \frac{1}{2}(\Gamma_{111} + \Gamma_{112} +
 \Gamma_{121} + \Gamma_{122} - \Gamma_{211} -
 \Gamma_{212} - \Gamma_{221} - \Gamma_{222}),
\nonumber\\
\Gamma_{Ro} &=& \frac{1}{2}(\Gamma_{111} + \Gamma_{121} +
 \Gamma_{211} + \Gamma_{221} - \Gamma_{112} - \Gamma_{122} -
 \Gamma_{212} - \Gamma_{222}),\nonumber\\
\Gamma_F &=& \frac{1}{2}(\Gamma_{111} + \Gamma_{121} 
+\Gamma_{212} + \Gamma_{222} - \Gamma_{112} - \Gamma_{211} 
- \Gamma_{221} - \Gamma_{122})\label{11}, \nonumber \\
\Gamma_{Fi} &=& \frac{1}{2}(\Gamma_{111} + \Gamma_{122} 
+\Gamma_{211} + \Gamma_{222} - \Gamma_{121} - \Gamma_{112} 
- \Gamma_{221} - \Gamma_{122}),\nonumber \\
\Gamma_{Fo} &=& \frac{1}{2}(\Gamma_{111} + \Gamma_{112} 
+\Gamma_{221} + \Gamma_{222} - \Gamma_{121} - \Gamma_{112} 
- \Gamma_{212} - \Gamma_{122}),\nonumber \\
\Gamma_E &=& \frac{1}{2}(\Gamma_{111} + \Gamma_{122} 
+\Gamma_{221} + \Gamma_{212} - \Gamma_{121} - \Gamma_{112} 
- \Gamma_{211} - \Gamma_{222}).\label{xxx}
\end{eqnarray}
Inverting these expressions we obtain a decomposition of the vertex function that is 
analogous to~(\ref{decompD1}) for the propagator:
 \begin{eqnarray}
   4\,\Gamma &=& \Gamma_R {1 \choose 1} {1\choose -1} {1\choose 1}
               + \Gamma_{Ri} {1 \choose -1}{1\choose 1} {1 \choose 1}
           + \Gamma_{Ro} {1 \choose 1} {1\choose 1} {1 \choose -1}\nonumber \\
             &+& \Gamma_F {1\choose -1} {1\choose 1}{1\choose -1}
           + \Gamma_{Fi} {1\choose 1}{1\choose -1}{1\choose -1}
             + \Gamma_{Fo} {1\choose -1}{1\choose -1}{1\choose 1}
  \label{eq: decompver}\\
           &+& \Gamma_E {1\choose -1}{1\choose -1}{1\choose -1}.\nonumber
 \end{eqnarray}

\subsection{Contracting Keldysh Vertices}

When performing calculations, we proceed according to 
the rules used in \cite{PeterH,mu}.  Bare vertices carry a factor of $\tau_3$ (the third Pauli 
matrix) because of the fact that a vertex of type-2 fields changes sign relative to a vertex 
of type-1 fields. When contracting indices, we must distinguish between internal and external 
indices.  For external indices, the product of the two column vectors carrying the same index 
is 
defined to be another column vector whose upper (lower) component 
is given by the product of upper (lower) components of the original
vectors:
 \begin{equation}
   {x_1 \choose x_2} {x_3 \choose x_4} = { x_1 x_3 \choose x_2 x_4} \, .
 \end{equation}
When contracting internal indices, the product of two column vectors is defined to be a scalar,
\begin{equation}
 {x_1 \choose x_2} {x_3 \choose x_4} = x_1 x_3 + x_2 x_4.
\end{equation}
The usefulness of this representation of the Keldysh formalism becomes clear when these 
contractions are done. Most of the huge number of terms that are produced in the real 
time formalism give zero, and these terms can be easily identified before any actual 
calculations are done.

\section{Bare Propagators}
In this section we use bare propagators so that~(\ref{eq: KMSD}) is satisfied, with 
the equilibrium distribution functions replaced by the appropriate non-equilibrium 
distributions.    

\subsection{Self Energy}

 We want to obtain integral expressions for the retarded and advanced electron self 
energies in non-equilibrium QED in the HTL limit. 

$\, $ From Fig. 1A we have, 
\begin{equation}
[\Sigma_{\lambda\alpha}]_{ab} = i\int \frac{d^4 S}{(2\pi)^4} 
(-ie\gamma^\mu_{\beta\alpha})(-i g_{\mu\nu} [\tau_3 D(S)\tau_3]_{ba})
(-ie \gamma^\nu_{\lambda \gamma}) i (P+S)_\tau \gamma^\tau _{\gamma\beta}[D(P+S)]_{ab}.
\end{equation}
In this expression both the self-energy and the propagators carry two Keldysh indices 
$\{a,b\}$ each of which takes values $\{1,2\}$.  
Doing the contraction over gamma matrices gives,
\begin{equation}
\Sigma_{ab} = 2ie^2\int \frac{d^4 S}{(2\pi)^4} (P \sla + S \sla) [\tau_3 D(S) \tau_3]_{ba} 
[D(P+S)]_{ab}.
\end{equation}
We contract the Keldysh indices using the rules discussed above and obtain a four component 
tensor  whose six terms (two propagators, each with three terms coming from the division 
into retarded, advanced and symmetric parts) can each be written as proportional to an 
outer product of column vectors of the form
\begin{equation}
{x \choose y} {u\choose v},
\end{equation}
where $\{x,y,u,v\}$ all have values $\pm 1$.  From~(\ref{sigma}) and (\ref{sigmaf}) 
it is clear that for $\Sigma_R$, $\Sigma_{A}$ and $\Sigma_F$ the only non-zero contributions 
will come from the terms proportional to 
\begin{equation}
{1\choose -1} {1 \choose 1}\, ,\,\,\,\,\,\,\,\,\,\,{ 1\choose 1}{1\choose -1}
\, ,\,\,\,\,\,\,\,\,\,\, 
{1\choose -1} {1 \choose -1},
\end{equation}
respectively.   The results are,
\begin{eqnarray}
\label{dra1}
\Sigma_R(P) &=& ie^2\int \frac{d^4S}{(2\pi)^4} ( P \sla+ S \sla)[D_F(S) D_R(P+S) + 
D_A(S) D_F(P+S)], \nonumber \\
\Sigma_A(P) &=& ie^2\int \frac{d^4S}{(2\pi)^4} ( P \sla+ S \sla)[D_F(S) D_A(P+S) + 
D_R(S) D_F(P+S)], 
\label{sigma1} \\
\Sigma_F(P) &=& ie^2\int \frac{d^4S}{(2\pi)^4} ( P \sla+ S \sla)[D_F(S) D_F(P+S) 
+ D_A(S) D_R(P+S) + D_R(S) D_A(P+S)]. \nonumber 
\end{eqnarray}
We rewrite these expressions using the notation $P_1 = P+S$ and $r_1 = D_R(P_1)$, etc.  
In addition we use the HTL limit and write $P \sla+ S \sla \rightarrow S \sla$.  We obtain,
\begin{eqnarray}
\Sigma_R(P) &=& ie^2\int \frac{d^4S}{(2\pi)^4} S \sla (f_S r_1 + a_S f_1),\nonumber \\
\Sigma_A(P) &=& ie^2\int \frac{d^4S}{(2\pi)^4} S \sla (f_S a_1 + r_S f_1), \label{10} \\
\Sigma_F(P) &=& ie^2\int \frac{d^4S}{(2\pi)^4} S \sla (f_S f_1 + a_S r_1 + r_S a_1). \nonumber 
\end{eqnarray}

\subsection{Vertex}
We consider the 1PI three-point function in QED. The integral for this vertex is given by 
the following expression (Fig. 2).
\begin{eqnarray}
(\Gamma^{\alpha\eta}_{\phi})(-Q,K,P)_{abc}& =& \int \frac{d^4 S}{(2\pi)^4} 
\{ (-ie\gamma^\gamma_{\alpha\beta})(-i g_{\mu\nu} [\tau_3 D(S)]_{ab})(-ie 
\gamma^\nu_{\theta\eta}) 
i(P \sla + S \sla)_{\tau\theta} \nonumber \\
&&[\tau_3 D(P+S)]_{bc}(-ie\gamma^\phi_{\lambda\tau}) i (Q \sla + S \sla)_{\beta\lambda}
[\tau_3 D(Q+S)]_{ca}\},
\end{eqnarray}
where $a,b,c$ are Keldysh indices.
After contracting gamma matrices we obtain,
\begin{equation}
(\Gamma^{\alpha\eta}_{\phi})_{abc} = 2e^3(\gamma^\lambda \gamma^\phi \gamma^\tau)_{\alpha\eta} 
\int \frac{d^4 S}{(2\pi)^4} (Q+S)_{\tau} (P+S)_\lambda [\tau_3 D(S)]_{ab} [\tau_3 D(P+S) ]_{bc}
[\tau_3 D(Q+S)]_{ca}.
\end{equation}
We use the contraction rules discussed previously.  We obtain an eight component tensor for 
the vertex function whose nine terms (three propagators, each with three terms coming from 
the division into retarded, advanced and symmetric parts) can each be written as proportional 
to an outer product of column vectors of the form
\begin{equation}
{x \choose y} {u\choose v} {w\choose z},
\end{equation}
where $\{x,y,u,v,w,z\}$ all have values $\pm 1$.  From~(\ref{11}) it is clear that for 
${\Gamma}_R$, ${\Gamma}_{Ri}$ and ${\Gamma}_{Ro}$ the only non-zero contributions will 
come from the terms proportional to 
\begin{equation}
{1\choose 1} {1\choose -1} {1 \choose 1}, \,\,\,\,\,\,\,\,\,\,\,{1\choose -1} {1\choose 1} 
{1 \choose 1}, \,\,\,\,\,\,\,\,\,\,\,
{1\choose 1} {1\choose 1} {1 \choose -1},
\end{equation}
respectively.  Similarly, for $\Gamma_F$, $\Gamma_{Fi}$, $\Gamma_{Fo}$ we need terms 
proportional to 
\begin{equation}
{1\choose -1} {1\choose 1} {1 \choose -1},\,\,\,\,\,\,\,\,\,\,\,{1\choose 1} {1\choose -1} 
{1 \choose -1},\,\,\,\,\,\,\,\,\,\,\,
{1\choose -1} {1\choose -1} {1 \choose 1},
\end{equation}
respectively. We will not need an expression for $\Gamma_E$.  
It does not appear in the integral expression for the one loop self energy or the one loop  
vertex in the second order effective theory, and therefore it is not needed to verify the 
Ward identity.  

The results are the following,

\begin{eqnarray}
(\Gamma_{\alpha\eta}^{\phi})_R(-Q,K,P) &=& e^3 
(\gamma^\lambda \gamma^\phi \gamma^\tau)_{\alpha\eta} 
\int \frac{d^4 S}{(2\pi)^4} (Q+S)_{\tau} (P+S)_{\lambda}
[r_S f_1 r_3 + f_S a_1 r_3 + a_S a_1 f_3],
\nonumber\\ 
(\Gamma_{\alpha\eta}^{\phi})_{Ri}(-Q,K,P) &=& e^3 
(\gamma^\lambda \gamma^\phi \gamma^\tau)_{\alpha\eta} \int \frac{d^4 S}{(2\pi)^4} 
(Q+S)_{\tau} (P+S)_{\lambda}[r_S r_1 f_3 + f_S a_1 a_3 + r_S f_1 a_3], \nonumber \\
(\Gamma_{\alpha\eta}^{\phi})_{Ro}(-Q,K,P) &=& e^3 
(\gamma^\lambda \gamma^\phi \gamma^\tau)_{\alpha\eta} \int \frac{d^4 S}{(2\pi)^4} 
(Q+S)_{\tau} (P+S)_{\lambda}[f_S r_1 r_3 + a_S f_1 a_3 + a_S r_1 f_3], \label{3p}\\
(\Gamma_{\alpha\eta}^{\phi})_F(-Q,K,P) &=& e^3 
(\gamma^\lambda \gamma^\phi \gamma^\tau)_{\alpha\eta} \int \frac{d^4 S}{(2\pi)^4} 
(Q+S)_{\tau} (P+S)_{\lambda}[r_S a_1 a_3 + a_S r_1 r_3 + f_S f_1 a_3 + f_S r_1 f_3 ],  
\nonumber\\ 
(\Gamma_{\alpha\eta}^{\phi})_{Fi}(-Q,K,P) &=& e^3 
(\gamma^\lambda \gamma^\phi \gamma^\tau)_{\alpha\eta} \int \frac{d^4 S}{(2\pi)^4} 
(Q+S)_{\tau} (P+S)_{\lambda}[ r_S a_1 r_3 + a_S r_1 a_3 + f_S f_1 r_3 + a_S f_1 f_3], 
\nonumber\\ 
(\Gamma_{\alpha\eta}^{\phi})_{Fo}(-Q,K,P) &=& e^3 
(\gamma^\lambda \gamma^\phi \gamma^\tau)_{\alpha\eta} \int \frac{d^4 S}{(2\pi)^4} 
(Q+S)_{\tau} (P+S)_{\lambda}[r_S r_1 a_3 + a_S a_1 r_3 + r_S f_1 f_3 + f_S a_1 f_3], 
\nonumber
\end{eqnarray}
where we have used the notation,
$P_1=P+S,\,\,\,\,P_3=Q+S$.    
We take the hard thermal loop limit $S>>P,Q$ and rewrite,  
\begin{equation}
(\gamma^\lambda \gamma^\phi \gamma^\tau) (Q+S)_{\tau} (P+S)_{\lambda} \rightarrow  
S_\tau S_\lambda (\gamma^\lambda \gamma^\phi \gamma^\tau) = 2S^\phi S \sla - 
\gamma^\phi S^2 \rightarrow 2S^\phi S \sla,
\end{equation}
where we have used the fact that terms in the integrand containing the factor 
$\gamma^\phi S^2$ do not contribute to leading order in the HTL approximation, 
as can be seen from power counting \cite{ref1}.

\subsection{Verification of the Ward Identities}

In order to verify the Ward identity between the three-point 
vertex function and the self-energy, we need to split propagators in 
the integral expressions for the vertex functions.  Using~(\ref{RA}) 
we obtain the identities, 
\begin{equation}
D_{R/A}(P_1) D_{R/A}(P_3) = \frac{1}{P_3^2 - P_1^2}(D_{R/A}(P_1) - D_{R/A}(P_3)).
\label{split}
\end{equation}

%\begin{eqnarray}
%D_{R/A}(P_1) D_{R/A}(P_3) &=& \frac{1}{P_3^2 - P_1^2}(D_{R/A}(P_1) - D_{R/A}(P_3)),\nonumber \\
%
%D_F(P_1) D_{R/A}(P_3) &=& \frac{1}{P_3^2 - P_1^2}D_F(P_1), \label{split} \\
%
%D_{R/A}(P_1) D_F(P_3) &=& -\frac{1}{P_3^2 - P_1^2} D_F(P_3) \, .\nonumber 
%\end{eqnarray}

We rewrite the integral expressions for the vertices~(\ref{3p}) by splitting $D(P_1)$ and 
$D(P_3)$ using~(\ref{split}).  We obtain,

\begin{eqnarray}
\Gamma^\phi_{R}(-Q,K,P) &=& e^3\int \frac{d^4 S}{(2\pi)^4} 
\frac{2S^\phi   S \sla } {(Q+S)^2 - (P+S)^2} 
[ r_S f_1 + f_S (a_1 - r_3) -a_S f_3],
\nonumber \\
\Gamma^\phi_{Ri}(-Q,K,P) &=& e^3\int \frac{d^4 S}{(2\pi)^4} 
\frac{2S^\phi  S \sla } {(Q+S)^2 - (P+S)^2} 
[ -r_S f_3 + f_S (a_1 - a_3) + r_S f_1],
\nonumber \\
\Gamma^\phi_{Ro}(-Q,K,P) &=& e^3\int \frac{d^4 S}{(2\pi)^4} 
\frac{2S^\phi  S \sla } {(Q+S)^2 - (P+S)^2} 
[ f_S(r_1-r_3) + a_S f_1 - a_S f_3],
\label{15}\\
\Gamma^\phi_{F}(-Q,K,P) &=& e^3\int \frac{d^4 S}{(2\pi)^4} 
\frac{2S^\phi  S \sla } {(Q+S)^2 - (P+S)^2} 
[ r_S(a_1-a_3) + a_S(r_1-r_3) + f_S f_1 - f_S f_3],
\nonumber \\
\Gamma^\phi_{Fi}(-Q,K,P) &=& e^3\int \frac{d^4 S}{(2\pi)^4} 
\frac{2S^\phi  S \sla } {(Q+S)^2 - (P+S)^2} 
[r_S (a_1 - r_3) + a_S (r_1-a_3)+ f_S f_1 ],
\nonumber \\
\Gamma^\phi_{Fo}(-Q,K,P) &=& e^3\int \frac{d^4 S}{(2\pi)^4} 
\frac{2S^\phi  S \sla } {(Q+S)^2 - (P+S)^2} 
[r_S (r_1 - a_3) + a_S (a_1 - r_3) - f_S f_3].
\nonumber
\end{eqnarray}
  Using the HTL approximation, we can write 
$[(Q+S)^2 - (P+S)^2] \rightarrow 2 S\cdot(Q-P)$.   We contract~(\ref{15}) 
with $K_\phi = (Q-P)_\phi$ and drop terms that have poles on the same side of the 
real axis in the $s_0$ plane (i.e., terms proportional to $a_S a_1$, $r_S r_3$, etc.), 
since they give zero after the contour integration is performed.  We obtain,
\begin{eqnarray}
K_\phi \Gamma^\phi_{R}(-Q,K,P)& =& e^3 \int \frac{d^4 S}{(2\pi)^4} S \sla
[f_S(r_1-r_3) + a_S f_1 - a_S f_3],
\nonumber \\
K_\phi \Gamma^\phi_{Ri}(-Q,K,P)& =& e^3 \int \frac{d^4 S}{(2\pi)^4} S \sla
[ -r_S f_3 + f_S (a_1 - a_3) + r_S f_1],
\nonumber \\
K_\phi \Gamma^\phi_{Ro}(-Q,K,P)& =& e^3 \int \frac{d^4 S}{(2\pi)^4} S \sla
[ r_S f_1 + f_S (a_1 - r_3) -a_S f_3],\label{contract} \\
K_\phi \Gamma^\phi_{F}(-Q,K,P)& =& e^3 \int \frac{d^4 S}{(2\pi)^4} S \sla
[  r_S(a_1-a_3) + a_S(r_1-r_3) + f_S f_1 - f_S f_3],\nonumber \\
K_\phi \Gamma^\phi_{Fi}(-Q,K,P)& =& e^3 \int \frac{d^4 S}{(2\pi)^4} S \sla
[r_S (a_1 - r_3) + a_S (r_1-a_3) + f_S f_1 ],\nonumber \\
K_\phi \Gamma^\phi_{Fo}(-Q,K,P)& =& e^3 \int \frac{d^4 S}{(2\pi)^4} S \sla
[ r_S (r_1 - a_3) + a_S (a_1 - r_3) - f_S f_3].\nonumber
\end{eqnarray}

Using~(\ref{10}) these expressions can be written,
\begin{eqnarray}
K_\phi \Gamma^\phi_{R}(-Q,K,P)& = & -ie(\Sigma_A(P) - \Sigma_R(Q)),\nonumber \\
K_\phi \Gamma^\phi_{Ri}(-Q,K,P) &=& -ie(\Sigma_A(P) - \Sigma_A(Q)),\nonumber \\
K_\phi \Gamma^\phi_{Ro}(-Q,K,P)& = &-ie(\Sigma_R(P) - \Sigma_R(Q)),\label{wd}
\\
K_\phi \Gamma^\phi_{F}(-Q,K,P)& = & -ie(\Sigma_F(P) - \Sigma_F(Q)),\nonumber \\
K_\phi \Gamma^\phi_{Fi}(-Q,K,P)& = & -ie\Sigma_F(P), \nonumber \\
K_\phi \Gamma^\phi_{Fo}(-Q,K,P)& = & ie\Sigma_F(Q),\nonumber 
\end{eqnarray}
which verifies that the Ward identities are obeyed by QED HTLs in non-equilibrium.  
We notice that the
Ward identities in non-equilibrium are structurely identical to those in
equilibrium.  

\section{Resummed Propagators}

In this section we will consider a type of higher order effective theory.  We will show that an identity which has the same form as the Ward identity is satisfied for a certain set of diagrams that form part of a second order effective theory.  We note that this relation is not really a Ward identity since it does not relate all diagrams contributing at a given order to the Green functions under consideration.\footnote{To calculate 
amplitudes with ultra-soft external energy scales ($\sim g^2$ times the hard momentum scale) consistently to any given 
order in perturbation theory, some additional type of resummation must be done, beyond the 
HTL resummation.  It is not known how such a resummation could be performed; there is no 
simple self-consistent prescription like the HTL formalism.   A second order effective theory 
of the type discussed above would include some of the diagrams that should be resummed, but 
it would 
not produce a self-consistent perturbation theory for amplitudes with ultra-soft external momenta.  In this sense, there is no direct analogy to the HTL effective theory which 
produces a self-consistent perturbation theory for amplitudes with soft external lines.} 
We will work with the propagator shown in Fig. 3.   The counterterm in this figure is included so that the HTL resummation provides a resummation of the perturbative expansion without changing the original Lagrangian.  Diagramatically, the counterterm is necessary to avoid the double counting of the HTL [Fig. (1A)].  
Naively, it appears that the pinch singularities which are regulated in the HTL propagator could re-emerge in this propagator.  We have shown however, that as with the 
HTL effective theory, such a second order effective propagator represents a resummation of all terms containing 
pinch singularities, 
and therefore we expect that the singularity is regulated in the same way as in the case of the HTL propagator \cite{mhm}.  
 The corresponding vertex function is given in Fig. 4.  As before, the counterterm is necessary to avoid double counting.    

We use the superscript $*$ to indicate a HTL resummed propagator 
(and not complex conjugation).  The resummed retarded and advanced propagators are given by (Fig. 1), 
\begin{eqnarray}
G_R^* =G_R + G_R \Sigma_R G_R^*, \nonumber \\
G_A^* = G_A + G_A \Sigma_A G_A^*, \label{GRA}
\end{eqnarray}
where $\Sigma_{R/A}$ is the HTL self-energy (as calculated in Section IIIA for fermions).  
The resummed symmetric propagator is given by \cite{mhm}, 
\begin{eqnarray}
G_F^*(P) &=& N(P)[G_R^*(P) - G_A^*(P)] + \{\Sigma_F(P) - N(P)[\Sigma_R(P) - \Sigma_A(P)]\} G^*_R(P) G^*A(P) \nonumber \\
&=& \Sigma_F(P) G_R^*(P) G_A^*(P). \label{resumprop}
\end{eqnarray}
The effective vertex is given by [Fig. 2],
\begin{eqnarray}
(\Gamma_\mu^*)_{\alpha\beta} = -ie(\gamma_\mu)_{\alpha\beta} + (\Gamma_\mu)_{\alpha\beta},
\label{effver}
\end{eqnarray}
where $(\Gamma_\mu)_{\alpha\beta}$ is the HTL vertex, as calculated in Section IIIB.

\subsection{The Self-Energy}

We begin by calculating the self-energy shown in Fig. 3A.  The result is,
\begin{eqnarray}
\Sigma^{*}_R(P) &=& -\frac{ie^2}{2}\int \frac{d^4S}{(2\pi)^4} \gamma^\mu (f^*_S \tilde{r}^*_1 + a^*_S \tilde{f}^*_1) \gamma_\mu,\nonumber \\
\Sigma^{*}_A(P) &=& -\frac{ie^2}{2}\int \frac{d^4S}{(2\pi)^4} \gamma^\mu(f^*_S \tilde{a}^*_1 + r^*_S \tilde{f}^*_1) \gamma_\mu,\label{101} 
\end{eqnarray}

%%%%%%%%%
\subsection{The Vertex}

We calculate the vertex shown in Fig. 4.  The integral is given by,
\begin{eqnarray}
(\Gamma^\phi)^{**}_{\alpha\eta}(-Q,K,P)_{cba} = && -ie(\gamma^\phi)_{\alpha\eta} + \int \frac{d^4 S}{(2\pi)^4}  
(-ie\gamma^\mu)_{\alpha\lambda} 
iS^*_{\lambda\theta}(Q+S)_{ec}  \\ 
&&(\Gamma^\phi)^*_{\theta\tau}(-Q-S,K,P+S)_{ebd} iS^*_{\tau\beta}(P+S)_{ad} (-ie\gamma^\nu)_{\beta\eta}(-ig_{\mu\nu}\tau_3   D^*(S)\tau_3)_{ca}.\nonumber
\end{eqnarray}
We use the tensor representations of the propagator and vertex as before, and do the 
contractions over Keldysh indices to obtain e.g. $(\Gamma^\phi_{Ri})^{**}_{\alpha\eta}(-Q,K,P)$.  
Suppressing  momentum variables  and Dirac indices we obtain,
\begin{eqnarray}
{\Gamma^\phi}^{ **}_{Ri} = -ie\gamma^\phi -\frac{i}{2} e^2\int \frac{d^4S}{(2\pi)^4} \gamma^\mu 
[\tilde{a}^*_3 {\Gamma^\phi}^*_F \tilde{r}^*_1 r^*_S + \tilde{f}^*_3 
{\Gamma^\phi}^*_{Ro} \tilde {r}^*_1 r^*_S + \tilde {a}^*_3 {\Gamma^\phi}^*_{Ri}
 \tilde{f}^*_1 r^*_S + \tilde{a}^*_3 {\Gamma^\phi}^*_{Ri} \tilde{a}^*_1 
f^*_S] \gamma_\mu .\label{rsG}
\end{eqnarray}
To verify the identity we contract this expression with $K_\phi$.  We use the 
Ward identities derived in Section IIIB for the HTL vertices $\Gamma$ to obtain 
expressions for the effective vertices $\Gamma^*$. From~(\ref{wd}) and~(\ref{effver}) 
we obtain,
\begin{eqnarray}
K_\phi \Gamma^{\phi *}_{Ri}(-Q-S,K,P+S) &=& ie(\tilde a^{*-1}_1 - \tilde a_3^{*-1}),\nonumber \\
K_\phi \Gamma^{\phi *}_{Ro}(-Q-S,K,P+S)& = &ie(\tilde r^{*-1}_1 - \tilde r_3^{*-1}),\label{wd2}
\\
K_\phi \Gamma^{\phi *}_{F}(-Q-S,K,P+S)& = & -ie(\Sigma_F(P_1)- \Sigma_F(P_3))\nonumber 
\end{eqnarray}
where $\tilde{a}_1^{*-1} = P \sla + S \sla - \Sigma_A(P+S)$, etc.  Note that~{(\ref{xxx}) shows that $\Gamma_F$ is zero at the tree level. 
Contracting~(\ref{rsG}) with $K_\phi$ and using~(\ref{resumprop}),~(\ref{101}),
and~(\ref{wd2}) we obtain 
\begin{eqnarray}
K_\phi \Gamma^{\phi **}_{Ri}(-Q,K,P) = ie(\tilde{a}_P^{**-1} - \tilde{a}_Q^{**-1}),
\label{FR}  
\end{eqnarray}
where $\tilde{a}_P^{**-1} = P \sla - \Sigma_A^{*}(P)$.
Equation~(\ref{FR}) has the same form as~(\ref{wd2}).  
Similar relations hold for the other components of 
${\Gamma^\phi}^{**}$.

\section{Conclusions} 
  
In this paper we have studied the retarded and advanced electron self-energies and the 
1PI three-point vertex in QED.  We have used the Keldysh representation of the RTF of 
finite temperature field theory.  We have worked within the HTL approximation.  We have 
shown that, out of equilibrium, the standard Ward identities are obeyed by the HTL 
self-energy and three-point vertex.  
The  analysis suggests that one could use the same procedure to show that the Ward 
identities are preserved out of equilibrium for higher n-point functions, in the HTL 
limit.  This result implies that the HTL effective action is gauge invariant both in 
and out of equilibrium. 

At next order we have shown that the same identity is obeyed by a set of one loop 
self-energy and one loop three-point functions which are constructed 
from HTL effective propagators and vertices.  The result holds both in and out of 
equilibrium and suggests that non-equilibrium calculations can be carried out in 
analogy to equilibrium ones.

\vspace*{1cm}

\centerline{\bf ACKNOWLEDGMENTS}

We would like to thank R. Kobes for helpful discussions.

\begin{figure}
\centerline{\psfig{figure=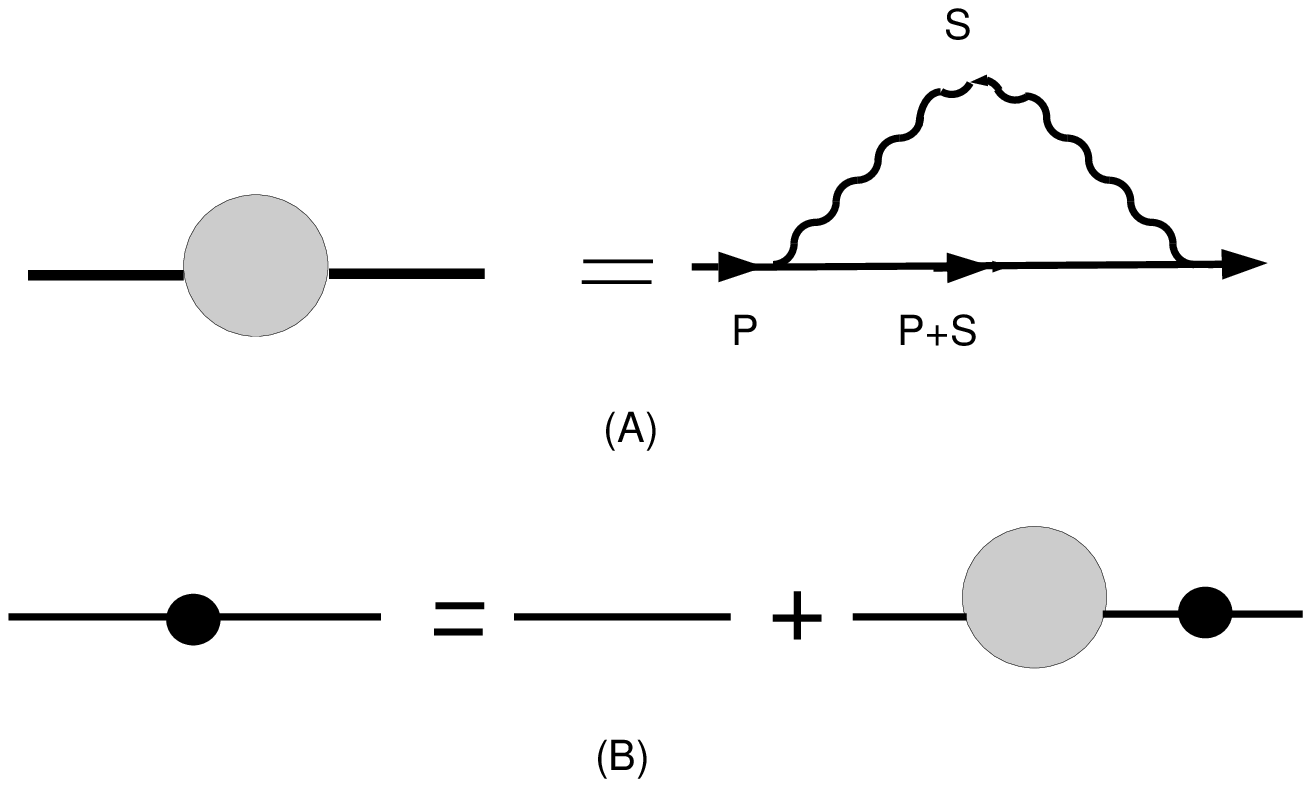,width=10cm}}
\caption{(A) HTL electron self energy with hard loop momenta;
(B) Schwinger-Dyson equation for the HTL effective fermion propagator.}
 \end{figure}
\begin{figure}
\centerline{\psfig{figure=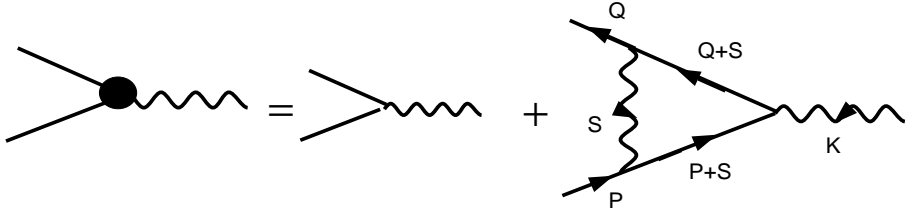,width=12cm}}
\caption{HTL effective vertex with hard loop momenta.}
\end{figure}

\newpage

\begin{figure}
\centerline{\psfig{figure=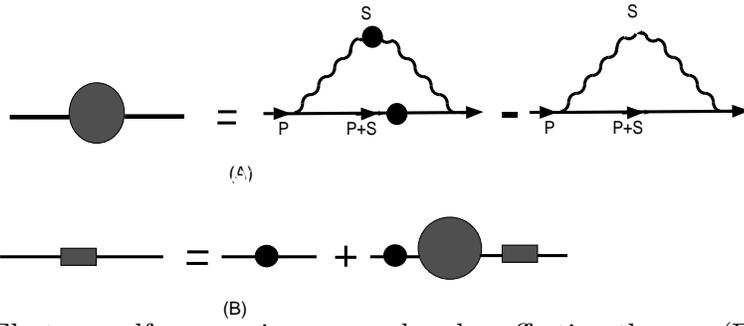,width=10cm}}
\caption{ (A)  Electron self energy in a second order effective theory;
(B) Schwinger-Dyson equation for the fermion propagator in a second order
effective theory.}
 \end{figure}

\begin{figure}
\centerline{\psfig{figure=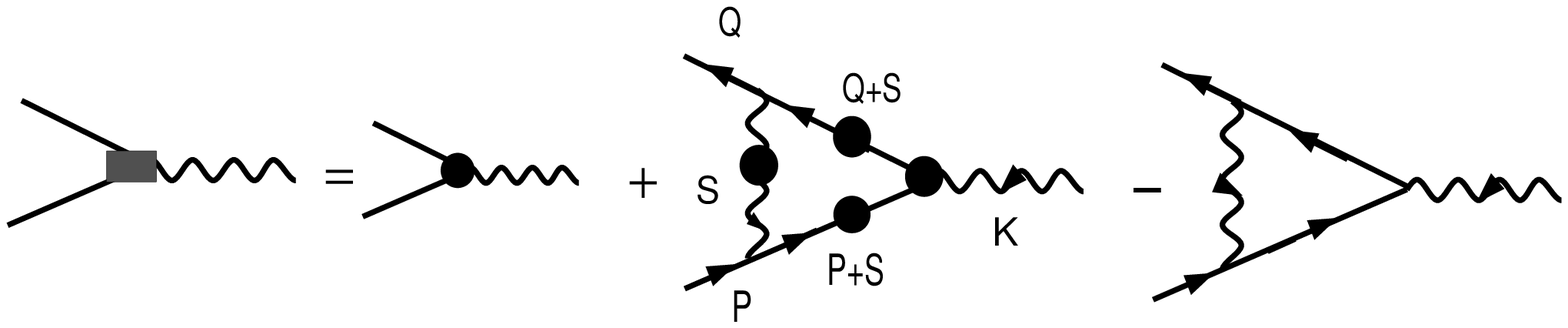,width=12cm}}
\caption{Effective vertex in a second order theory.}
\end{figure}


\begin{references}
\bibitem{ref1} E. Braaten and R.D. Pisarski, Nucl. Phys. {\bf B337}, 569
(1990).
\bibitem{ref2} M.H. Thoma, in {\it Quark-Gluon Plasma 2}, edited by R. Hwa
(World Scientific, Singapore, 1995), p.51.
\bibitem{taylor1} J. Frenkel and J.C. Taylor, Nucl.Phys. {\bf B374}, 156 
(1992).
\bibitem{BP92} E. Braaten and R.D. Pisarski, Phys. Rev. D {\bf 45}, R1827 
(1992).
\bibitem{ref6} K. Chou, Z. Su, B. Hao, and L. Yu, Phys. Rep. {\bf 118}, 1
(1985).
\bibitem{ref7} L.V. Keldysh, JETP {\bf 20}, 1018 (1965).
\bibitem{mhm} M.E. Carrington, Hou Defu, and M.H.Thoma, hep-ph/9708363.
\bibitem{PeterH} P.A. Henning, Phys. Rep. {\bf 253}, 235 (1995).
\bibitem{mu} M.E. Carrington and U. Heinz, Eur. Phys. J. C {\bf 1}, 619 
(1998).

\end{references}
\end{document}